\documentclass[reprint,amsmath,amssymb,aps,superscriptaddress
]{revtex4-2}
\usepackage{graphicx}
\usepackage{dcolumn}
\usepackage{hyperref}
\usepackage[mathlines]{lineno}
\usepackage{bbold}
\usepackage{tabularx}
\usepackage{cleveref}
\usepackage{physics}
\usepackage{siunitx}
\usepackage{multirow}
\usepackage{subcaption}
\usepackage{caption}
\usepackage{booktabs}

\begin{document}

\title{A Quantum Leaky Integrate-and-Fire Spiking Neuron and Network}

\author{Dean Brand}
\email{dean.brand@nithecs.ac.za}
\affiliation{Department of Physics and School of Data Science and Computational Thinking, Stellenbosch University, Stellenbosch, 7604, South Africa}
\affiliation{National Institute for Theoretical and Computational Sciences (NITheCS), Stellenbosch, 7604, South Africa}
\author{Francesco Petruccione}
\affiliation{Department of Physics and School of Data Science and Computational Thinking, Stellenbosch University, Stellenbosch, 7604, South Africa}
\affiliation{National Institute for Theoretical and Computational Sciences (NITheCS), Stellenbosch, 7604, South Africa}

\date{\today}

\begin{abstract}
    Quantum machine learning is in a period of rapid development and discovery, however it still lacks the resources and diversity of computational models of its classical complement. With the growing difficulties of classical models requiring extreme hardware and power solutions, and quantum models being limited by noisy intermediate-scale quantum (NISQ) hardware, there is an emerging opportunity to solve both problems together. Here we introduce a new software model for quantum neuromorphic computing --- a quantum leaky integrate-and-fire (QLIF) neuron, implemented as a compact high-fidelity quantum circuit, requiring only 2 rotation gates and no CNOT gates. We use these neurons as building blocks in the construction of a quantum spiking neural network (QSNN), and a quantum spiking convolutional neural network (QSCNN), as the first of their kind. We apply these models to the MNIST, Fashion-MNIST, and KMNIST datasets for a full comparison with other classical and quantum models. We find that the proposed models perform competitively, with comparative accuracy, with efficient scaling and fast computation in classical simulation as well as on quantum devices.
\end{abstract}

\maketitle

\section{Introduction}
\label{sec:Introduction}

In the development of modern computers and machine learning models, there is a rapidly approaching ceiling of ability which will necessarily require a paradigm shift in information processing. As the end of Moore's law looms \cite{waldrop_chips_2016}, there is a growing need to turn to alternative computational models which are able to overcome the scaling issues of computationally hard and expensive models, alongside a gargantuan ocean of data \cite{hashem_rise_2015}. The solutions to these problems need to be coupled across hardware and software.

A promising and demonstrated potential solution to the processing of high-dimensional and complicated data lies in the field of quantum computing, which already has sophisticated approaches to optimisation and learning. Quantum computing, specifically Quantum Machine Learning (QML) \cite{biamonte_quantum_2017,schuld_introduction_2015,tacchino_artificial_2019}, has shown rapid growth in recent years to catch up to the computational offering of its classical counterpart. With the additional benefit of being able to process and learn both classical and quantum data, the field is filling a niche ahead of its time. The most promising models for QML on classical data are based on variational quantum circuits \cite{peruzzo_variational_2014, cerezo_variational_2021}, which encode data into quantum systems to utilise the natural tendency of these systems to find optimal energetic states.

Quantum hardware, however, is still in the Noisy Intermediate Scale Quantum (NISQ) era of devices \cite{preskill_quantum_2018,cheng_noisy_2023}, which significantly limits the capabilities of the software models present. The hardware limits the software by being too susceptible to noise from the open quantum systems within the device to stay in a coherent state for long enough for the completion of the complicated and deep quantum circuits \cite{breuer_theory_2010}.

Another approach to the ceiling of computational ability we currently face, comes from the original inspiration of machine intelligence. Artificial neurons are meant to mimic the behaviour of information processing and transfer that occurs in organic brains. However, the modern implementation of artificial neural networks (ANNs) is based on a very simple model which reduces the complexity of a brain to nodes and connections as an abstraction of neurons and synapses. Although modern ANNs have seen incredible success in performing computations and learning tasks, they are not a perfect model, which is shown by comparison to human brains which they aspire to replicate. For example, in the modern large language model (LLM) GPT-3, there are 175 billion parameters which consume an estimated $190\,000,\si{\kWh}$ to train \cite{anthony_carbontracker_2020,budennyy_eco2ai_2022,eshraghian_training_2023}, compared to the human brain consisting of 100 billion neurons and operating on the scale of $20\,\si{\watt}$ \cite{markovic_physics_2020}. There are caveats to this comparison but the scale is still evident of poor efficiency of ANNs.

This gives rise to the need for biologically plausible models which are computationally efficient, more like organic brains \cite{hodgkin_quantitative_1952}. There is also a significant hardware limitation leading to this inefficiency, broadly referred to as the von Neumann Bottleneck, caused by the computing architecture of the same name proposed 80 years ago \cite{zou_breaking_2021}. The hardware problem is proposed to be solved by neuromorphic processors, which are the foundation for more efficient software models \cite{benjamin_neurogrid_2014,merolla_million_2014,davies_loihi_2018}.

The most promising evolution of ANNs to the neuromorphic computing paradigm are spiking neural networks (SNNs) \cite{maass_networks_1997, roy_towards_2019}, which use binary activation pulses of current that encode temporal information in addition to intensity. In contrast ANNs represent only mean spike firing rates, compressing the temporal dimension and losing that information \cite{haykin_neural_2009}. These models offer significantly more information density at a much greater efficiency than ANNs on conventional von Neumann hardware.

Within the collection of SNN models, there are several artificial neuron models which offer varying levels and trade-offs of complexity, efficiency, and biological plausibility. A sweet-spot choice for SNNs which compete with ANNs is that of the leaky integrate-and-fire (LIF) neuron, first studied in 1907 using biological systems \cite{brunel_lapicques_2007, burkitt_review_2006}. This neuron is a popular choice for highly performant SNNs, due to its balance of simplicity and efficiency while being able to handle temporal data better than its ANN counterpart \cite{yamazaki_spiking_2022,wang_supervised_2020}.

With both the quantum and neuromorphic models of computing and machine learning now introduced, as well as both of their recent advancements, it sets a natural stage for research of their intersection. There has already been novel proposals of quantum neuromorphic hardware \cite{pfeiffer_quantum_2016,salmilehto_quantum_2017,prati_quantum_2017,sanz_invited_2018,guo_y_-m_quantum_2021,markovic_quantum_2020} and software \cite{kristensen_artificial_2021,chen_accelerating_2022}, however not of the form of adapting a spiking neuron to a gate-based quantum computer as we do here.

The rest of the paper is structured as follows, in \Cref{sec:Neuron} we derive the QLIF model from the classical LIF dynamics and discuss its utility. This is followed by \Cref{sec:Neural_Network} where we connect the QLIF neurons in a network and compare its performance to competitive quantum, spiking, and classical models. In \Cref{sec:Discussion} we discuss and analyse the results of the network performance in the context of a hybrid quantum neuromorphic setting, and finally in \Cref{sec:Conclusion} we provide concluding remarks and an outlook on future work.

\section{Neuron}
\label{sec:Neuron}

The dynamics of the biological LIF neuron can be modeled as a low-pass filter resistor-capacitor (RC) circuit \cite{hodgkin_quantitative_1952,eshraghian_training_2023}. This circuit aggregates the membrane potential, $U(t)$, as it is driven by input current spikes and over time with no stimulus decays to equilibrium. This is expressed as
\begin{figure}[t!]
    \centering
    \captionsetup{justification=raggedright}
    \includegraphics[scale=0.75]{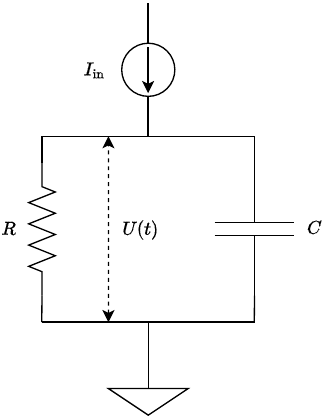}
    \caption{A low-pass filter resistor-capacitor (RC) circuit of the LIF neuron. The circuit describes how membrane potential dynamics of a neuron are simulated by a time-varying input current $I_\mathrm{in}$. The membrane potential, $U(t)$, is decomposed into a resistor and capacitor which form a linear differential equation describing a spiking neuron.}
    \label{fig:lif_circuit}
\end{figure}
\begin{equation}
\label{eq:rc_differential}
    \tau\dv{U(t)}{t} = -U(t) + I_\mathrm{in}R,
\end{equation}
with $\tau = RC$ as a time constant for the dynamics of the circuit. The LIF circuit is represented in \Cref{fig:lif_circuit}. Solving this differential equation and approximating with the forward Euler method gives a time-discretised expression for the membrane potential as
\begin{equation}
\label{eq:lif_discrete}
    U[t] = \beta U[t-1] + (1-\beta)I_\mathrm{in}[t],
\end{equation}
where the decay rate is represented by $\beta = \exp(-1/\tau)$. The `fire' part of the LIF neuron comes from the feature that once the membrane potential has been driven enough to reach a threshold value, $U_\mathrm{thr}$, the neuron fires an output spike of current by expelling the built-up membrane potential. To alter the form of the neuron expression to be applicable to deep learning tasks, the input current term can be seen as a weighted contribution of a binary input spike, $X[t]$, and simplified as $I_\mathrm{in}[t] = WX[t]$ for a learnable weight $W$. This simplification, along with the fire-reset feature, lead to the expression
\begin{equation}
\label{eq:lif_final}
    U[t] = \underbrace{\beta U[t-1]}_\mathrm{decay} + \underbrace{WX[t]}_\mathrm{input} - \underbrace{S_\mathrm{out}[t-1]U_\mathrm{thr}}_\mathrm{reset},
\end{equation}
where the output spike is generated under the condition
\begin{equation}
\label{eq:lif_fire_condition}
    S_\mathrm{out}[t] = \begin{cases}
            1,\quad\text{if } U[t]>\theta \\
            0,\quad\text{otherwise}.
    \end{cases}
\end{equation}
\begin{figure}[t!]
    \centering
    \captionsetup{justification=raggedright}
    \includegraphics[width=\linewidth]{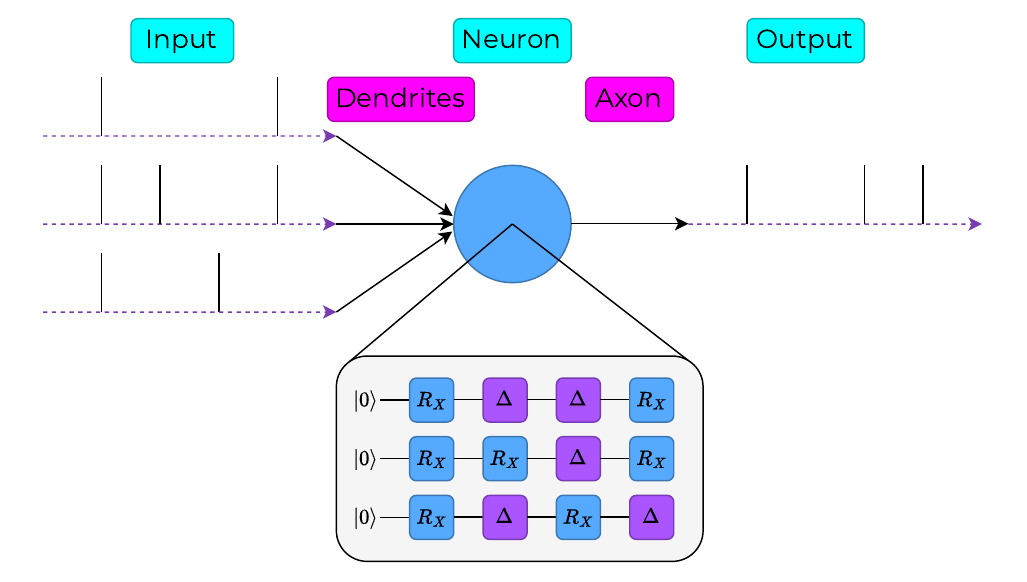}
    \caption{A Quantum Leaky Integrate-and-Fire (QLIF) neuron processing input spike stimuli. Spikes to the excited state population are modelled through rotation gates ($R_X$). The lack of a spike is processed as a delay gate ($\Delta$), during which the qubit does nothing for a time $t$, and the excited state population decays exponentially. In the construction of a QLIF neural network, multiple inputs are aggregated through the dendrites to produce a set of outputs after being processed by the neuron.}
    \label{fig:qsn_input_output}
\end{figure}

Now, to recreate this behaviour on a quantum computer, to create a QLIF neuron, we begin with the spiking behaviour. For an input stimulus, the simplest way to encode a spike in a quantum circuit is a simple rotation gate, by an angle corresponding to a variable input intensity, $\theta$, which takes the qubit from the ground state, $\ket{0}$, to some excited state $R_X(\theta)\ket{0}$. This leads to a direct spike in the excited state population, which is the mechanism corresponding to a membrane potential in the classical model. The threshold mechanism works identically in both cases, however in the quantum model it is limited to a range of $0$ to $1$.

For the exponential leak behaviour in the case of no input spike, the quantum model shows its strength of utilising environmental noise. In the open quantum system of a quantum circuit, when an excited state is left to evolve through time then it undergoes simple $T_1$ relaxation to decay to the ground state \cite{brand_markovian_2024}. This $T_1$ relaxation process can be approximated by an exponential decay function, which is exactly what is required for the leak of the neuron.

For an input spike there is a rotation gate to spike the excited state population, and for no spike there is a `delay' gate during which the qubit does nothing and naturally interacts with the environment to decay to the ground state. This process is depicted in \Cref{fig:qsn_input_output}.

Now, after each input spike, or lack thereof, there needs to be a measurement of the circuit to see the excited state population and if it has passed the firing threshold or not. If it is above the threshold, then the state is reset to the ground state $\ket{0}$ and an output spike is recorded. If it is below the threshold, then the next spike needs to be processed.

However, once the circuit is measured, it needs to be constructed again for the next spike to be processed. This means that the previous state of the qubit would need to be reconstructed, by creating the previous circuit each time iterating with the newest spike processing gate of a rotation or delay. This would be very inefficient and create unnecessarily deep circuits which also do not account for the thresholding mechanism.

\begin{figure}[t!]
    \centering
    \captionsetup{justification=raggedright}
    \includegraphics[width=0.8\linewidth]{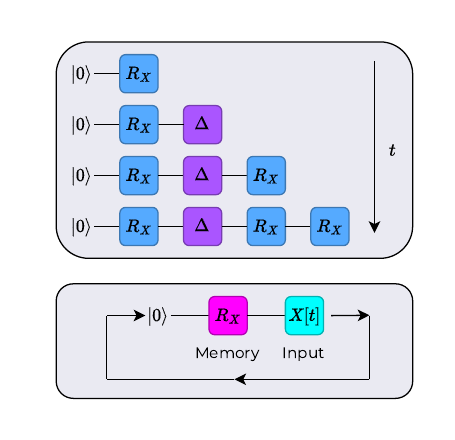}
    \caption{Compact circuit structure of a QLIF neuron processing binary spike input stimulus. Input spikes are processed as rotation gates ($R_X$) to increase the excited state population. Absent input spikes are processed as delay gates ($\Delta$), during which the qubit does nothing for a short period of time and is decayed by noise from the environment. Due to the simple circuit structure, the previous state can be compacted to a single rotation gate which can re-instate the excited state population of the previous measurement. This serves as the memory of the neuron, before the next input is processed.}
    \label{fig:qlif_recurrent}
\end{figure}

Fortunately, because the circuits only consist of $R_X$ rotations, the previous excited state population can be reinstated by a rotation gate of an angle that would correspond to the previous state before measurement. As can be seen in \Cref{fig:qlif_recurrent}, this leads to very compact 2-gate single-qubit circuits to process each spike, which can then naturally be processed in parallel over more qubits, and stay decoupled.

To create an expression for the time-discretised behaviour of the QLIF neuron similar to that of \Cref{eq:lif_discrete}, we consider how the excited state population, $\alpha$, is influenced by each spike. First, the `memory' contribution needs to be considered as an action that reinstates the previous excited state population such that the next input can be processed. To reinstate the population, an $R_X$ gate can be applied with an angle of
\begin{equation}
    \label{eq:qlif_memory}
    \varphi[t] = 2\arcsin\left(\sqrt{\alpha[t]}\right).
\end{equation}

For an input spike, corresponding to an $R_X(\theta)$ gate, leads to a measured state population of
\begin{equation}
    \label{eq:qlif_Theta}
    \Theta[t] = \sin^2\left(\frac{(\theta + \varphi[t])X[t]}{2}\right),
\end{equation}
for
\begin{equation}
\label{eq:X_spikes}
    X[t] = \begin{cases}
        1, \quad\text{spike,} \\
        0, \quad\text{no spike.}
    \end{cases}
\end{equation}
To model the exponential decay from noise and the $T_1$ relaxation process, we can make some simplifications to the approach of open quantum systems and Markovian modelling \cite{brand_markovian_2024}. In the simple case of an excited state relaxing to the ground state, the relaxation is an exponential decay based on the characteristic $T_1$ time of the device, which has the form $\propto \exp(-\tau/T_1)$ for some time period $\tau$. In real quantum hardware this phenomenon is easy to achieve with the natural noise present, and the time parameter $\tau$ can easily be varied for the desired decay rates.
\begin{figure}[t!]
    \centering
    \captionsetup{justification=raggedright}
    \includegraphics[width=\linewidth]{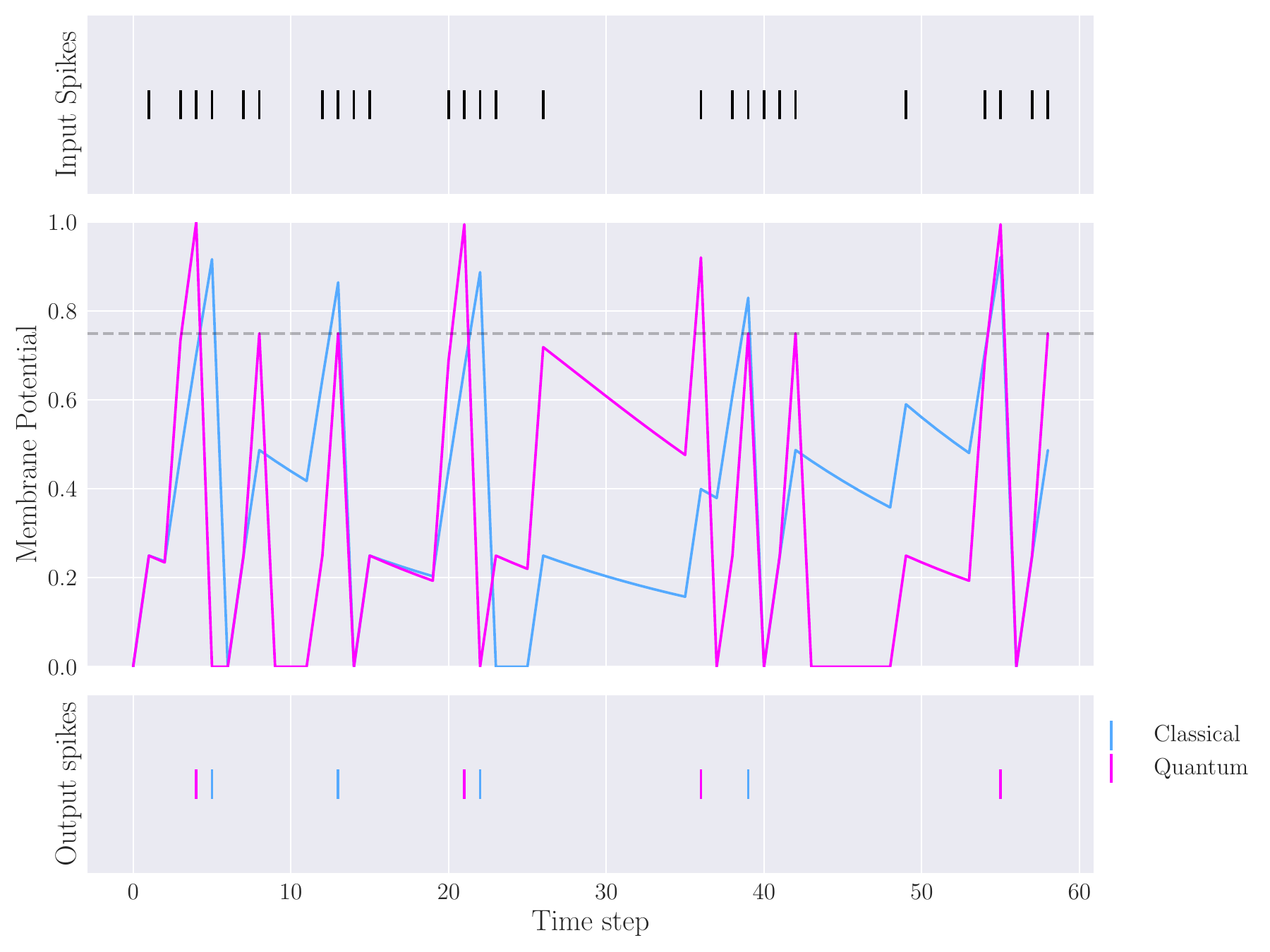}
    \caption{Classical LIF vs QLIF neuron behaviour processing a train of input spikes, producing a series of output spikes based on a threshold shown by the grey dashed line. Input spikes are randomly generated here, and processed over 60 time-steps by the LIF and QLIF neurons to demonstrate the behavioural similarity between the neurons. Classical LIF membrane potential is shown in blue, while the QLIF excited state population is shown in magenta.}
    \label{fig:lif_qlif}
\end{figure}

However, for simulation and modelling of this behaviour, we can consider how to implement this decay effect manually. Consider that the exponential decay gradually decreases the excited state population, after it has been increased by an $R_X$ gate. This decay can also be mimicked by a rotation in the opposite direction, by an angle corresponding to how much the state would have decayed. This angle is given by
\begin{equation}
    \label{eq:gamma_angle}
    \gamma[t] = -2\arcsin\left(\sqrt{\alpha[t] \exp\left(-\frac{\tau}{T_1}\right)}\right),
\end{equation}
after which it can take the same form of contribution to the excited state population as \Cref{eq:qlif_Theta}. Thus, as an expression for the dynamics of the excited state population, as a function of variables $\theta$ and $\tau$, we have
\begin{equation}
    \label{eq:qlif_main}
    \begin{aligned}
    \alpha[t+1] &= \sin^2\left( \frac{\left(\theta + \varphi[t]\right) X[t]}{2} \right) \\
    &\quad+ \sin^2\left( \frac{\left(\gamma[t] + \varphi[t]\right)(X[t] - 1)}{2} \right).
    \end{aligned}
\end{equation}

In this form, the inputs and outputs can easily be vectorised, scaling as $\mathcal{O}(n)$, as compared to the simulation of other quantum circuits such as variational models, which scale as $\mathcal{O}(n^2)$ for $n$ qubits \cite{schuld_machine_2021}. Furthermore, due to the QLIF model only requiring 2 gates and a single qubit for each evaluation, it eliminates all concern of noise disrupting the results, as is the case in deeper circuits of previous QML models.

On the contrary, the small levels of noise present in the circuit are necessary for the functionality of the QLIF neuron, making it an ideal fit for NISQ era models of QML. This formulation of the QLIF neuron also closely mimics the classical LIF model, as can be seen in \Cref{fig:lif_qlif}, where the form of the spikes and decay are the same, down to variable tweaking to achieve identical behaviour.

\section{Neural Network}
\label{sec:Neural_Network}

To construct a network of these new neurons and perform more complex computational tasks, it is a straightforward approach to have them serve as a drop-in replacement for classical ANN neurons, identical to how the classical SNN is designed. In this drop-in replacement, the input is converted initially to a spike-rate, with other options of encoding available \cite{borst_information_1999,panzeri_sensory_2010,qiang_yu_rapid_2013}. This spike rate is used to generate randomly distributed spike trains, typically sampled from a Poisson distribution \cite{kempter_spikebased_1998}, for all of the first layer neurons.

\begin{figure}
    \centering
    \captionsetup{justification=raggedright}
    \includegraphics[width=\linewidth]{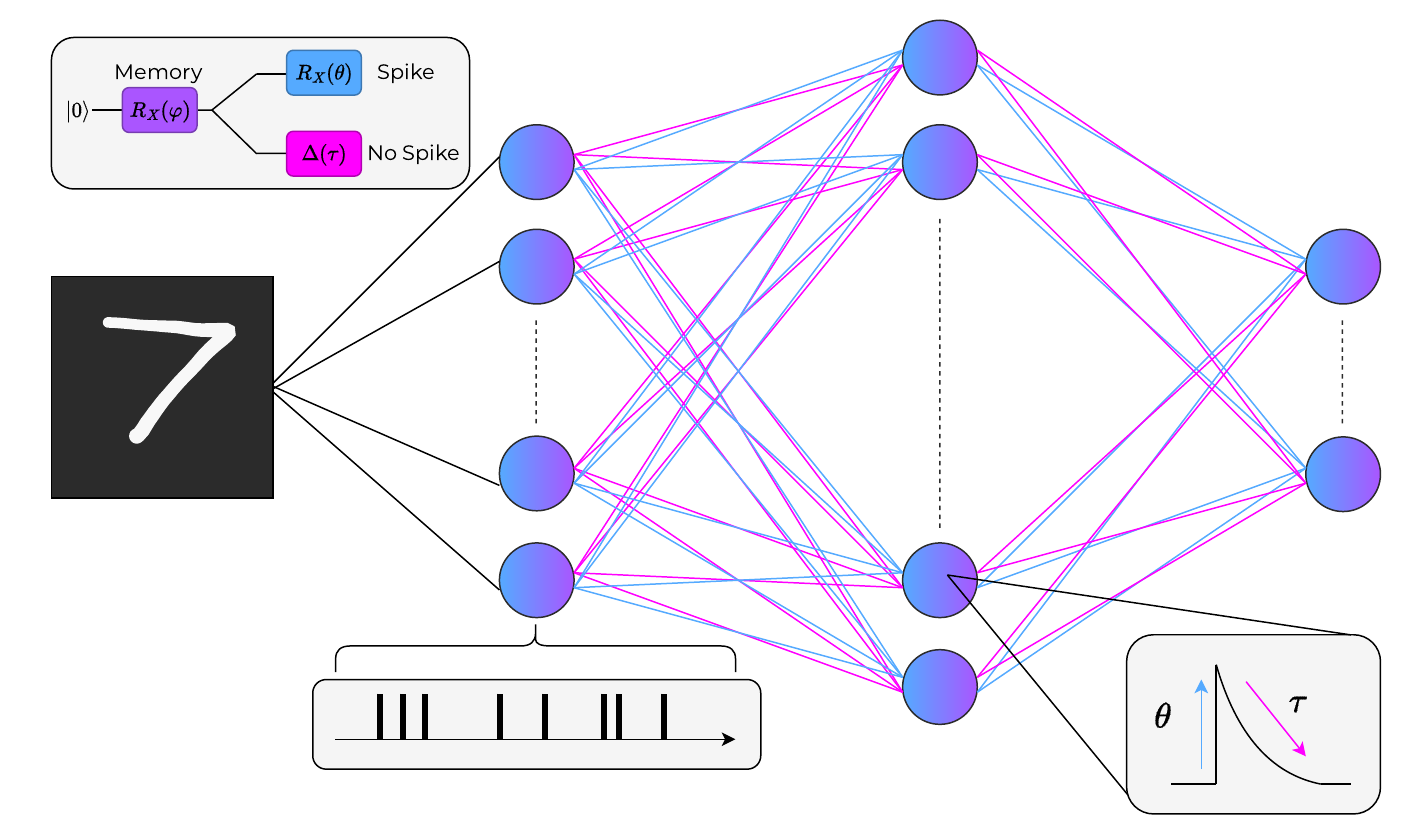}
    \caption{A Quantum Spiking Neural Network structure, processing an image converted to brightness values and then to spike trains to pass through the network. The network trains two planes of weighted synapses ($\theta$ and $\tau$) in parallel for each neuron. The parameter $\theta$ controls the spike intensity through the angle of the $R_X$ rotation gate. The parameter $\tau$ controls the decay rate through the delay gate $\Delta$, during which the qubit is left to idle and undergo relaxation to a ground state (Bottom right). Only one of these operations occurs based on the binary input, after a memory rotation gate which reinstates the excited state of the previous time step measurement (Top left). Each neuron processes a series of input spikes/no-spikes (Bottom left). Throughout the diagram $\theta$ is blue and $\tau$ is magenta.}
    \label{fig:seven_qsnn}
\end{figure}

The first layer processes the first wave of spikes in parallel and propagates the output to the next layer through the weighted synapse connections. Contrary to classical SNNs, however, the Quantum Spiking Neural Network (QSNN) model has two parameters to train: the spike intensity (rotation angle $\theta$) and the decay rate (delay time $\tau$), as opposed to just the weighted input $WX[t]$ from \Cref{eq:lif_final}. This means that the two weights are trained in parallel, for the processing of each wave of spikes in the forward pass through the layers of the fully connected neural network, as shown in \Cref{fig:seven_qsnn}.

At the end of the forward pass, the output layers can be configured to be decoded in whichever way is required to have a meaningful output. For example, in a classification task, the output can be a layer of $N$ neurons for $N$ classes, with the neuron that has the highest spiking rate being the predicted class.

For the backward pass, there are a multitude of learning rules and training methods available to choose from as with most classical ML models \cite{wang_supervised_2020,yamazaki_spiking_2022,malcolm_comprehensive_2023,eshraghian_training_2023}. In the model presented here, we have chosen a gradient descent methodology and backpropagation using spikes to train the network. Specifically, we use the backpropagation through time (BPTT) method and the SpikeProp algorithm \cite{bohte_error-backpropagation_2002} for temporally encoded SNNs.

To calculate the gradients to propagate through the network using this method, one needs to consider not only the direct gradients of all of the neurons as in ANNs, as the \textit{immediate influence} to the network, but also all previous time steps where the neurons have been stimulated as a \textit{prior influence} to the network. As the gradient of the loss function with respect to the weights being trained, this can be generally expressed as
\begin{equation}
    \label{eq:bptt_chain}
    \frac{\partial\mathcal{L}}{\partial W} = \sum_t \frac{\partial\mathcal{L}[t]}{\partial W} = \sum_t \sum_{s\leq t} \frac{\partial\mathcal{L}[t]}{\partial W[s]},
\end{equation}
for all steps $s$ up to and including the latest time step $t$. For the calculation of the loss at each time step, the gradient can be expanded using the chain rule for all of the connected variables used in the network. For the weights used in the QSNN, $\theta$ and $\tau$ (here generalised as $W$), the gradients of the loss are expressed as follows,
\begin{equation}
    \label{eq:theta_gradient}
    \frac{\partial\mathcal{L}}{\partial W} = \frac{\partial\mathcal{L}}{\partial S} \frac{\partial S}{\partial \alpha} \frac{\partial \alpha}{\partial W},
\end{equation}
where $S$ is the output spike, and $\alpha$ is the excited state population from \Cref{eq:qlif_main}. Unfortunately, the QSNN also suffers the problem of dead neurons as in classical SNNs, where the output spike $S$ is a step function with respect to the excited state population, which is non-differentiable and thereby makes the training of the network impossible in this form. Fortunately, an elegant solution exists in the form of \textit{surrogate gradients} \cite{neftci_surrogate_2019,cramer_surrogate_2022}, during which the forward pass of the network is done as per usual, with a useful trick. In the backward pass the non-differentiable gradient is replaced with a function that mimics similar behaviour to the output spike but in a way that is differentiable and enables meaningful training. There are several candidates for a good surrogate gradient function, but in the network proposed here we use the arctan surrogate gradient,
\begin{equation}
    \label{eq:arctan_surrogate}
    \frac{\partial S}{\partial\alpha} \leftarrow \frac{\partial\tilde{S}}{\partial\alpha} = \frac{1}{\pi} \frac{1}{1 + (\alpha[t]\pi)^2}.
\end{equation}
Now, we have all the parts necessary to construct the QSNN. It is important to note, however, that this construction is a very simple proposal just to show the functionality as a proof-of-concept, and it can be improved in many ways which are outside the scope of the work presented here.

\section{Discussion}
\label{sec:Discussion}

To evaluate the performance of the QSNN made of QLIF neurons, we constructed a basic neural network applied to the MNIST digits image classification problem \cite{li_deng_mnist_2012}, as well as the Fashion-MNIST \cite{xiao_fashion-mnist_2017} and Kuzushiji-MNIST \cite{clanuwat_deep_2018} datasets, for a thorough demonstration of its performance and predictive capability. To understand its relative performance and advantages compared to other neural network models, we also applied a classical ANN, SNN, and a quantum variational classifier to the same datasets.

This ensemble of classical and quantum models on 3 classification problems was replicated using convolutional neural network frameworks as well. All models were kept very simple in their construction, and as similar as possible in terms of architecture decisions to keep the comparison as clear as possible. The models were constructed and evaluated in Python, making use of the PennyLane package for quantum simulation and PyTorch for neural networks \cite{bergholm_pennylane_2022, paszke_pytorch_2019}.

For the standard fully-connected networks, the following architecture was used; A greyscale image of size $28\times28$ was normalised and flattened as an input layer of size 784, followed by a fully connected weight layer to the hidden layer of $1\,000$ neurons, once again fully connected to the output layer of 10 neurons used as the classifier. In the quantum variational classifier \cite{schuld_circuit-centric_2020}, the 784 input neurons were connected to a layer of 10 neurons which feed directly as input to a quantum circuit of 10 qubits, comprising an angle embedding layer and a set of strongly entangling layers to process the data, which was then connected to an output layer of 10 classifier neurons.

In the convolutional upgrade of the models, the greyscale image of $28\times28$ pixels was normalised and flattened again to a layer of 784 neurons. This is then connected to a $5\times5$ convolutional kernel with 12 filters, followed by a $2\times2$ max-pooling function, followed by a fully-connected layer that maps $1\,024$ neurons to 10 outputs. In the quantum case, the Quanvolutional model was used \cite{henderson_quanvolutional_2020}. For all of the networks, the same cross-entropy loss function was used, and all used the Adam optimiser \cite{kingma_adam_2017}.

All of the models were trained using the same classical computer, including the noisy and noiseless simulations of quantum models. This ensures consistency in the results, and training the models several times allowed for statistically rigorous results. For each dataset, the models used a training set of $60\,000$ images and a test set of $10\,000$ images, split in batches and trained until a satisfactory convergence was reached, in a maximum of 5 epochs. For the quantum models, the results were obtained using noiseless simulations, although we intend to investigate a noisy real hardware setting to further show the relative comparison of the quantum models and the QSNN in a future work. The results of the simulations are summarised in \Cref{tab:results_summary}.
\begin{table}[t!]
    \centering
    \captionsetup{justification=raggedright}
    \begin{tabular}{ccccc}
        \toprule
        \multirow{2}{3em}{Model} & \multicolumn{3}{c}{Accuracy (\%)} & \multirow{2}{2.5em}{Time} \\
        \cmidrule{2-4}
         & MNIST & FMNIST & KMNIST & \\
        \midrule
        ANN & 97.84 & 87.24 & 90.38 & 30s \\
        SNN & 95.84 & 81.73 & 90.02 & 2m26s \\
        QNN & 89.65 & 82.90 & 67.18 & 54m22s \\
        Noisy QNN & 86.82 & 80.68 & 59.18 & 5h27m37s \\
        \textbf{QSNN} & \textbf{88.25} & \textbf{75.25} & \textbf{60.36} & \textbf{2m14s} \\
        CNN & 96.87 & 81.19 & 81.98 & 1m48s \\
        SCNN & 97.75 & 65.81 & 82.06 & 5m49s \\
        Quanv & 92.46 & 83.99 & 71.83 & 6h27m8s \\
        Noisy Quanv & 92.35 & 82.87 & 70.58 & 31h26m6s \\
        \textbf{QSCNN} & \textbf{90.62} & \textbf{70.19} & \textbf{66.02} & \textbf{5m40s} \\
        \bottomrule
    \end{tabular}
    \caption{Summarised results of the ten neural network models, quantum and classical, convolutional and fully-connected, noisy and noiseless, on the three image classification datasets of MNIST, Fashion-MNIST, and Kuzushiji-MNIST. Accuracy measured on test set of $10\,000$ images for each dataset. Training time measured on test set of $60\,000$ images for each dataset. Models proposed here shown in bold.}
    \label{tab:results_summary}
\end{table}

\begin{figure}
    \centering
    \captionsetup{justification=raggedright}
    \includegraphics[width=\linewidth]{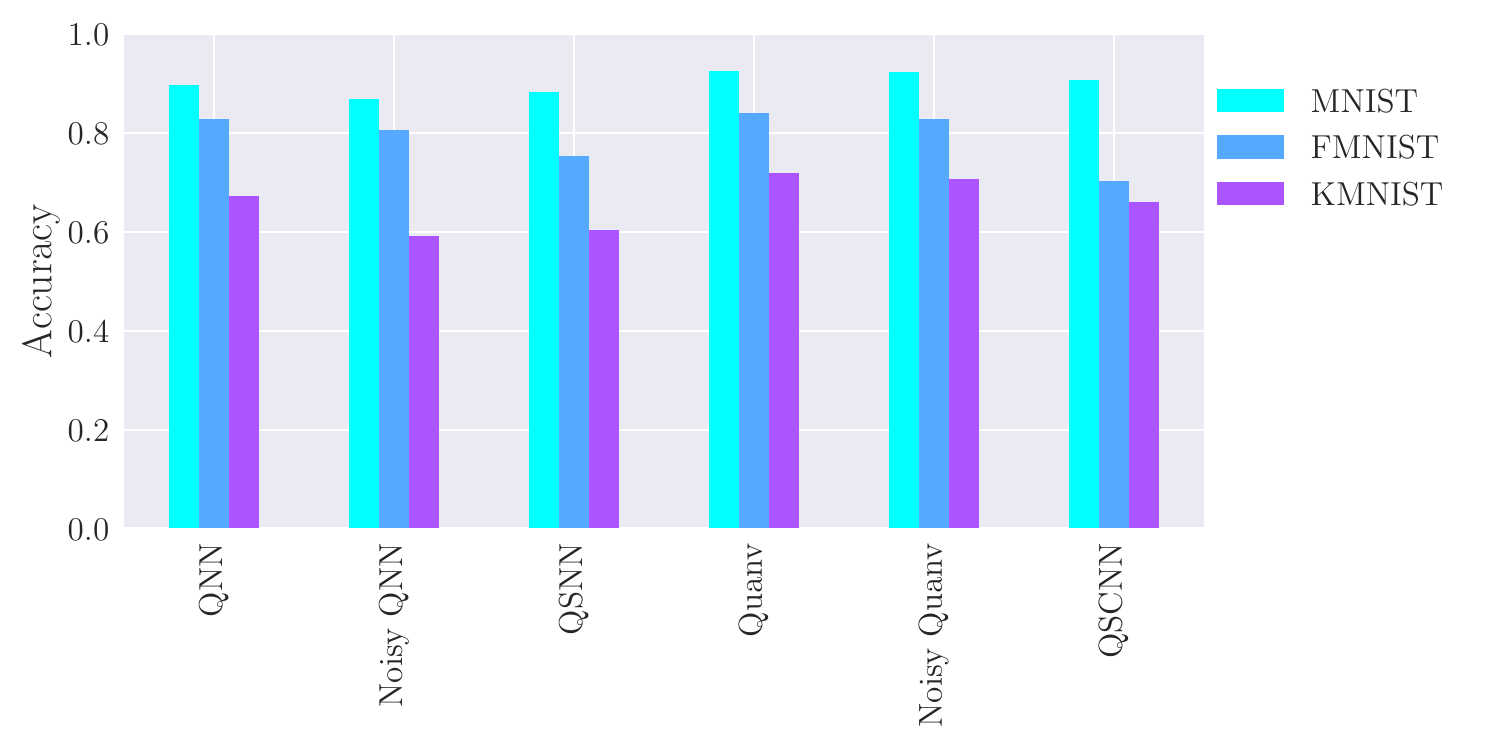}
    \caption{Accuracy comparison of six quantum neural network models, variational and spiking, convolutional and fully-connected, noisy and noiseless, on the three image classification datasets of MNIST, Fashion-MNIST, and Kuzushiji-MNIST. Accuracy measured on test set of $10\,000$ images for each dataset.}
    \label{fig:perf_acc}
\end{figure}

Upon deeper analysis of the results presented in \Cref{tab:results_summary}, it is clear to see where the advantage of the QSNN and QSCNN is. The accuracy across the datasets remains comparable with the other quantum models, but are 24 and 68 times, respectively, faster than their quantum competitors in a noiseless simulation. In a noisy simulation these factors extend to the quantum spiking models being 146 and 333 times, respectively, faster than quantum counterparts. These encouraging results can be seen in \Cref{fig:perf_acc}. For noisy simulations this training time disparity is exaggerated even more, with the QSNN and QSCNN results to be almost entirely unaffected by the noise present due to the minimalist QLIF architecture. These noisy simulations of the QNN and Quanv models are seen in \Cref{tab:results_summary} to have extreme training times relative to the newly proposed model, and suffer in accuracy, performing worse than the faster model in some cases. It is clear from this where the advantage and the promise of the newly proposed QSNN and QSCNN models are.

The results and metrics also indicate strongly that with a more suited NN and CNN architecture using the QLIF neuron, the performance can match and potentially exceed comparable models. These results are a simple benchmark for basic comparative performance, but should not be seen as the maximum potential of this new branch of QSNN. Additionally, the QLIF neuron has the advantage of being simple enough for mathematical simulation without the need for quantum computing frameworks or exclusively real hardware quantum devices. The mathematical flexibility and computational simplicity make this tool a powerful potential source of advantage and progress in the fields of QML and quantum neuromorphic computing.

\section{Conclusion}
\label{sec:Conclusion}

We have derived and proposed a new quantum spiking neuron, based on the leaky integrate-and-fire principles from classical neuromorphic computing. This QLIF neuron is highly efficient for a quantum circuit, making use of a maximum two rotation gates on a single qubit to process input binary spikes of data sequentially and recurrently. The neuron is not hindered by noisy quantum devices as most variational quantum circuits are, but rather requires the influence of noise to relax the excited states as a fundamental part of its functionality, thus being an ideal candidate for use on NISQ devices.

These QLIF neurons can be connected into fully connected as well as convolutional networks, for simple functionality and transposition from classical counterpart networks, to perform temporal data processing across many neuron connections on a quantum computer. These networks were shown to perform well and and train faster than variational quantum circuits on 3 benchmark datasets: MNIST, Fashion-MNIST, and Kuzushiji-MNIST. These benchmark results only scratch the surface of the capabilities of the QLIF neuron and the networks constructed using it. The work here presents a proof-of-concept of the idea of a quantum spiking neuron in this form which utilises natural quantum features to improve the efficiency of quantum neural networks and serve as a new avenue of research for quantum neuromorphic computing.

Many future works are planned to improve upon the computational models presented here, and to branch out into new computational models that utilise spiking neurons that have yet to be implemented in the realm of quantum computing. Such plans include improving network architecture to offer more flexibility in learning methodology and structure to suit each dataset with deeper and more complex neural networks. Additionally, these networks can be applied to specifically neuromorphic data, such as temporally recorded image data \cite{orchard_converting_2015,see_st-mnist_2020}, biological sensor data \cite{cramer_heidelberg_2022}, and time-series data such as video processing \cite{mueggler_event-camera_2017} or anomaly detection \cite{cherdo_time_2023}. Furthermore this computational model can be combined with quantum neuromorphic hardware models for a seamless pipeline of efficient data processing and learning on quantum devices \cite{pfeiffer_quantum_2016,salmilehto_quantum_2017,prati_quantum_2017,sanz_invited_2018,guo_y_-m_quantum_2021}.

With such an outlook and high performance of this computational model, the fields of quantum machine learning and quantum neuromorphic computing have a new realm of exploration to find quantum advantage and highly performant and efficient data processing, immediately applicable to industry and research endeavours.

\hspace{5pt}

\section*{Acknowledgments}
    This work is based on the research supported in part by the National Research Foundation of South Africa, Ref. PMDS22070532362. This work was funded by the South African Quantum Technology Initiative (SAQuTI) through the Department of Science and Innovation of South Africa. The funders played no role in study design, data collection, analysis and interpretation of data, or the writing of this manuscript. The authors acknowledge the Centre for High Performance Computing (CHPC), South Africa, for providing computational resources to this research project.

\section*{Code availability}
The data presented in this paper was created using Python, particularly the PennyLane and PyTorch packages. The code created to obtain and analyse these results is available on GitHub at \url{https://github.com/deanbrand/QSNN}.

\bibliographystyle{utphys}
\bibliography{references}
	
\end{document}